\documentclass[prb,reprint,aps,twocolumn]{revtex4-1}

\pdfoutput=1

\usepackage[utf8]{inputenc}
\usepackage[T1]{fontenc}
\usepackage{hyperref}
\usepackage{color}
\usepackage{graphicx}
\usepackage{units}
\usepackage{amsmath}
\usepackage[usenames,dvipsnames]{xcolor}

\hypersetup{colorlinks = true, citecolor = blue, breaklinks = true}

\begin{document}

\title{Strain-induced heteronuclear charge disproportionation in EuMnO$_3$}
\date{\today}
\author{Ulrich Aschauer}
\affiliation{Materials Theory, ETH Zurich, Wolfgang-Pauli-Strasse 27, CH-8093 Z\"urich, Switzerland}
\affiliation{Department of Chemistry and Biochemistry, University of Bern, Freiestrasse 3, CH-3012 Bern, Switzerland}
\author{Nathalie Vonr\"uti}
\affiliation{Materials Theory, ETH Zurich, Wolfgang-Pauli-Strasse 27, CH-8093 Z\"urich, Switzerland}
\affiliation{Department of Chemistry and Biochemistry, University of Bern, Freiestrasse 3, CH-3012 Bern, Switzerland}
\author{Nicola A. Spaldin}
\affiliation{Materials Theory, ETH Zurich, Wolfgang-Pauli-Strasse 27, CH-8093 Z\"urich, Switzerland}

\begin{abstract}
Charge disproportionation transitions in complex oxides most commonly link high-temperature phases containing identical cations with the same oxidation state and crystallographic site, to low-temperature phases in which charge transfer between these ions results in unequal oxidation states. Here we propose, based on density functional calculations, a related concept that we term \textit{heteronuclear} charge disproportionation, in which charge transfer occurs between different elements on different crystallographic sites. We show for the case of EuMnO$_3$ that such a transition from the experimentally observed Eu$^{3+}$Mn$^{3+}$O$_3$ phase to the so far unknown Eu$^{2+}$Mn$^{4+}$O$_3$ phase can be triggered by pressure and epitaxial strain. We then identify measurable signatures to aid in an experimental exploration of the complex pressure- and biaxial strain-dependent phase stability of EuMnO$_3$ that we hope to motivate with our predictions. We suggest other candidate crystal chemistries for heteronuclear charge disproportionation, in which novel physics could emerge from the coexistence of instabilities.
\end{abstract}

\maketitle

\section{Introduction}

Charge disproportionation occurs frequently in complex oxides, with examples including CaFeO$_3$ (in which formally 2Fe$^{4+}$ $\rightarrow$ Fe$^{3+}$ + Fe$^{5+}$) \cite{Woodward:2000id}, PbCrO$_3$ (3Cr$^{4+}$ $\rightarrow$ 2Cr$^{3+}$ + Cr$^{6+}$) \cite{Cheng:2015gu} and BaBiO$_3$ (Bi$^{4+}$ $\rightarrow$ Bi$^{3+}$ + Bi$^{5+}$) \cite{Lobo:1995ti} among many others\footnote{Note that of course in practice due to covalency the actual electron counts deviate from those implied by the formal charges}. It is often accompanied or followed by charge ordering, in which the ions of different charges arrange in a regular pattern on the crystal lattice, and it is associated with functionalities such as metal to insulator transitions \cite{Alonso:1999je}, ferroelectricity \cite{Alexe:2009iy} or the onset of magnetic ordering \cite{Woodward:2000id}. Usually, the transition occurs as a function of temperature, with the transition metal ions having equal partial charges in the high temperature phase, and different and integer charges, yielding insulating behavior, at low temperature. Importantly, charge ordering transitions can be affected by applied isostatic pressure \cite{Kawakami:2001cm, Cheng:2015gu}, or epitaxial strain \cite{Catalano:2014hu}, since the different ionic radii of the disproportionated ions cause a strong coupling between the charge state and the lattice. 

While conventional charge disproportionations involve ions of the same element on sites with the same high-temperature symmetry, recently an unusual charge ordering driven by charge transfer between ions on inequivalent lattice sites was demonstrated in perovskite-structure (HgMn$_3$)Mn$_4$O$_{12}$ \cite{Chen:2018hi}. At 490K the system undergoes a standard charge disproportionation transition, with the four formally Mn$^{3.25+}$ B-site cations disproportionating (but not ordering) to three Mn$^{3+}$ and one Mn$^{4+}$; the A-site Mn ions all have the same 3+ charge. At 240K there is a second transition in which an electron is transferred from the B site to the A site; both B- and A-site sublattices then charge order with two Mn$^{3+}$ and one Mn$^{4+}$ ions (B site), or two Mn$^{3+}$ and one Mn$^{2+}$ ions (A site) respectively. This {\it homonuclear} intersite charge transfer and charge ordering in turn breaks the inversion symmetry and induces a small ferroelectric polarization. 

\begin{figure}
\includegraphics[width=0.9\columnwidth]{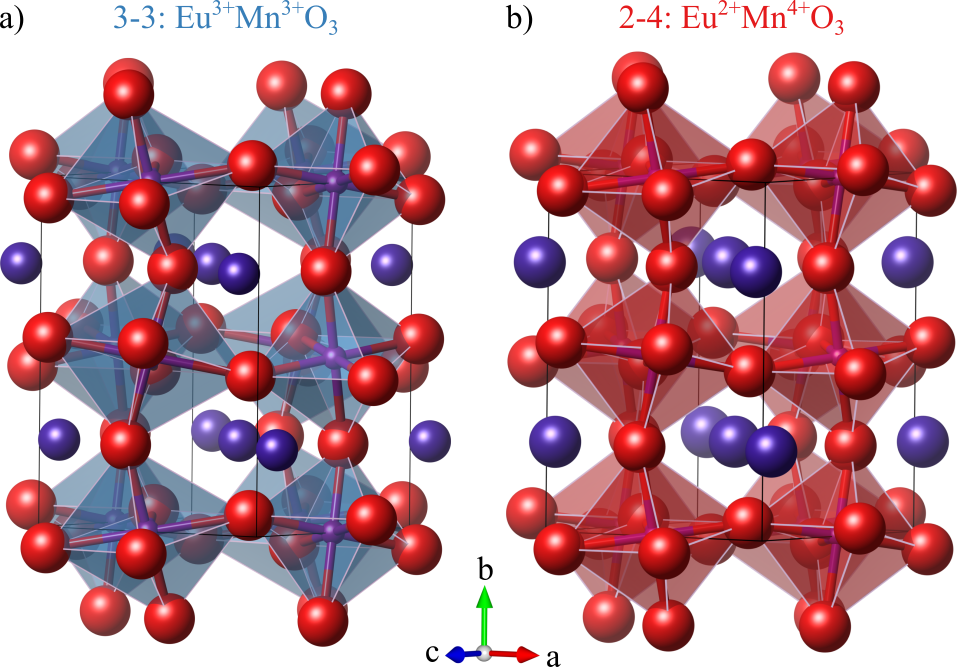}
\caption{\label{fig:structure}Calculated structures (this work) of a) the Eu$^{3+}$Mn$^{3+}$O$_3$ (3-3) states of perovskite EuMnO$_3$ and b) the proposed Eu$^{2+}$Mn$^{4+}$O$_3$ (2-4) state. Ions are drawn as spheres with sizes corresponding to their respective Shannon radii. Color code: Eu=blue, Mn=purple, O=red. Mn-O octahedra are shown in blue and red for the 3-3 and 2-4 phase respectively.}
\end{figure}

In this work we introduce the concept of {\it heteronuclear} intersite charge transfer and propose a charge disproportionation transition between two ions of different elements on different sites. We use the perovskite-structure oxide EuMnO$_3$ as our model material. Perovskites with both Eu$^{2+}$ and Eu$^{3+}$ oxidation states, which are compatible with the B-site transition metal adopting 4+ or 3+ oxidation states respectively, are known to exist; we refer to these as 2-4 and 3-3 phases in the following. Experimentally, EuMnO$_3$ exists in the 3-3 state at ambient conditions: It exhibits the cooperative Jahn-Teller distortion that is typical for Mn$^{3+}$ ions having d$^4$ occupation \cite{Choithrani:2009ua}, and strong octahedral rotations expected for the small tolerance factor (Table \ref{tab:radii}), $t=0.88$, of the 3-3 compound. Its 3-3 ground state is also consistent with those of other Eu perovskites of the mid-first-row transition metals, EuCrO$_3$ \cite{Taheri:2015cd}, EuFeO$_3$ \cite{Choquette:2015ko} and EuCoO$_3$ \cite{Pekinchak:kv}. The calculated structure of 3-3 EuMnO$_3$ is shown in Fig.~\ref{fig:structure}a.

In principle, however, EuMnO$_3$ can also occur as the Eu$^{2+}$Mn$^{4+}$O$_3$ 2-4 compound, whose calculated structure we show in Fig. \ref{fig:structure}b. While Eu$^{2+}$ A sites are typically restricted to early transition-metal B-site elements such as EuTiO$_3$ \cite{BussmannHolder:2015hc} and EuZrO$_3$ \cite{Saha:2016jm}, the prevalance of Mn$^{4+}$, for example in perovskites of the alkaline earth series (i.e. CaMnO$_3$ \cite{Poeppelmeier:1982ch}, SrMnO$_3$ \cite{Becher:2015df}, BaMnO$_3$ \cite{Rondinelli:2009tq}), suggests that the 2-4 state might be accessible.

\begin{table}
\caption{Shannon ionic radii (in \AA ) \cite{Shannon:1976vx} for the ions in the two possible states of EuMnO$_3$, leading to the listed tolerance
factors \cite{Goldschmidt:1926wn} and expected cubic lattice parameters. Also shown is the experimental pseudocubic lattice constant for the 3-3
phase, obtained from the cube root of the volume of a single formula unit.}
\begin{ruledtabular}
\begin{tabular}{l | l l}
     & 2-4                                  & 3-3 \\ \hline
Eu & 1.35 & 1.12 \\
Mn & 0.53 & 0.65 \\
O & 1.35 & 1.35 \\ \hline
Tolerance factor, $t$ & 1.02 & 0.88 \\
Cubic lattice $a = 2(r_\mathrm{Mn}+r_\mathrm{O})$ & 3.76 & 3.99 \\
Experimental pseudocubic lattice & & 3.85 \\
\end{tabular}
\end{ruledtabular}
\label{tab:radii}
\end{table}

The cubic lattice constants for EuMnO$_3$, determined from the ionic radii ($a=2(r_B+r_O)$), are 3.76 \AA\ and 3.99 \AA\ in the 2-4 and 3-3 phases respectively (Table \ref{tab:radii}). While the experimental pseudocubic lattice constant of the 3-3 phase (3.85 \AA, obtained from the cube root of the volume of one formula unit) is smaller than this value due to the strong octahedral rotations, it remains larger than that of the 2-4 phase. This suggests that positive pressure or compressive epitaxial strain should convert the 3-3 ground state to the 2-4 phase. Experimentally, however, existence of the 2-4 phase has not yet been reported in measurements performed up to 54 GPa \cite{Muthu:2012bs, Mota:2014dv}. This could be for a number of reasons: First, the transition could be kinetically hindered. Second, the properties of the 2-4 phase might be similar to that of the 3-3, so that its existence was overlooked. Or finally, the naive arguments based on ionic radii and tolerance factors might not capture the full behaviour so that the transition does not occur or occurs only at prohibitively high pressure.

Here we use density functional theory (DFT) calculations to clarify the existence and nature of such a 3-3 to 2-4 transition and to identify measurable signatures of its occurrence. We find that the heteronuclear charge disproportionated 2-4 phase is only slightly higher in energy than the 3-3 phase and has a smaller unit-cell volume. Therefore it should be accessible at reasonable values of pressure or compressive epitaxial strain. We analyse space groups, lattice parameters, magnetic moments and Raman spectra in the two phases to aid in the experimental exploration of the complex pressure-dependent phase stability of EuMnO$_3$ that we hope to motivate with our predictions. Finally, we suggest other materials containing two multivalent cations in which heteronuclear charge disproportionation should occur. 

\section{Computational Method}

We begin by calculating the theoretical structures for the 3-3 and 2-4 phases. For the 3-3 phase the known experimental \textit{Pnma} structure was used as the starting point for our calculations \cite{Choithrani:2009ua}. For the 2-4 phase we took the structure of the metastable perovskite SrMnO$_3$ \cite{Becher:2015df}, which has comparable ionic radii (Eu$^{2+}$ = 1.35 \AA\ and Sr$^{2+}$ = 1.36 \AA\ \cite{Shannon:1976vx}) and relaxes into an \textit{Imma} space group. In order to impose a specific oxidation state, we constrained the occupation matrix during structural relaxation until the corresponding lattice structure was converged \cite{Allen:2014ft}. The constraint was then released for a final relaxation. All calculations were performed using the VASP code \cite{Kresse:1993ty,Kresse:1994us,Kresse:1996vk,Kresse:1996vf} with the PBEsol exchange-correlation functional \cite{Perdew:2008fa} and the default VASP Eu PAW potential \cite{Blochl:1994uk,Kresse:1999wc} with \textit{f} electrons in the valence. The valence configurations of our PAW potentials were Eu(5s, 6s, 5p, 4f), Mn(4s, 3p, 3d) and O(2s, 2p). Wavefunctions were expanded in planewaves up to a kinetic energy of 700 eV and reciprocal space was sampled using a $\Gamma$-centred 6x4x6 mesh for the orthorhombic 20-atom cell with $b$ as the long axis. Internal coordinates and cell shapes/normal axes were relaxed until forces converged below $10^{-3}$ eV/\AA\ and stress below $5\cdot 10^{-5}$ eV/\AA$^3$. Phonon calculations were performed within the frozen phonon method as implemented in phonopy \cite{Togo:2008jt} and the Raman activity was computed via mode amplitude-dependent changes in the dielectric constants similarly to the approach implemented in Ref. \onlinecite{Fonari:2013}.

We applied a Dudarev \cite{Dudarev:1998vn} DFT+U correction to both the Eu $f$ and Mn $d$ states. It was previously shown that U$_\textrm{Mn}$ has an effect on the pressure dependence of the magnetic ground state \cite{Qiu:2017kl}, with U$_\textrm{Mn}$ values of 1 and 2 eV giving the same qualitative results but quantitatively different antiferromagnetic to ferromagnetic transition pressures. We found that U$_\textrm{Mn}$ also has a marked effect on the charge ground state, with U$_\textrm{Mn}$ = 1 and 3 eV resulting in the 2-4 and 3-3 ground states respectively. At U$_\textrm{Mn}$ = 2 eV the two charge states are nearly degenerate, with the 3-3 phase being slightly lower in energy (see appendix figure \ref{fig:struct_stab}). Our calculations therefore reproduce the experimentally known charge state for U$_\textrm{Mn}$ > 2 eV. Given that 3-3 manganites such as TbMnO$_3$ typically show an AFM to FM transition at U values between 2 and 3 eV \cite{Franchini:2012fla, Fedorova:2018aa}, we chose to carry out our calculations with U$_\textrm{Mn}$ of 2 eV to yield both the correct charge state and, according to the literature, the correct magnetic ground state. We emphasize, however, that the exact pressure at which our calculated transitions occur is highly sensitive to the choice of U$_\textrm{Mn}$, and so our calculated values should not be taken as precise predictions of the experimental transition points. 

The U value on the Eu \textit{f} states was set to 10 eV to position them in agreement with x-ray photoelectron spectroscopy (XPS) measurements for the 3-3 phase \cite{Kurmaev:1998ic}. We note that while with the Eu \textit{f} electrons in the core as in Ref. \onlinecite{Qiu:2017kl}, U$_\textrm{Mn}$=2 eV results in the correct A-AFM magnetic structure (see Figure \ref{fig:mag_stab}a), we find that with the \textit{f} electrons in the valence, FM alignment of the Mn moments is slightly favoured, independently of the value of U$_\textrm{Mn}$ and U$_\textrm{Eu}$ and the nature of the Eu magnetic order (see supporting Figure \ref{fig:mag_stab}b and c). This highlights the commonly observed delicate dependence of AFM $\rightarrow$ FM Mn transitions in the rare-earth manganites on the computational setup \cite{Franchini:2012fla}. In view of this shortcoming, and given that our interest here is in the charge ground state, we \textit{impose} G-type AFM Mn order for the 2-4 phase and A-type AFM Mn order for the 3-3 phase respectively \cite{Mukovskii:1998io} and do not address the recently predicted AFM to FM magnetic transition \cite{Qiu:2017kl}. The nature of the Eu ordering is not known experimentally; we find ferromagnetic layers of spin moments alternating antiferromagnetically along the long axis of the 20-atom unit-cell to have the lowest energy in the 3-3 state (by 3 meV per cell) and adopt this arrangement throughout this study, also for the 2-4 phase. We do not include spin-orbit coupling in our calculations, which we expect to result in a zero net magnetic moment state on the $f^6$ Eu$^{3+}$ ions ($S=3$, $L=3$ and $J=0$).

\section{Results \& Discussion}

\subsection{Isostatic pressure}

\begin{figure}
\includegraphics[width=\columnwidth]{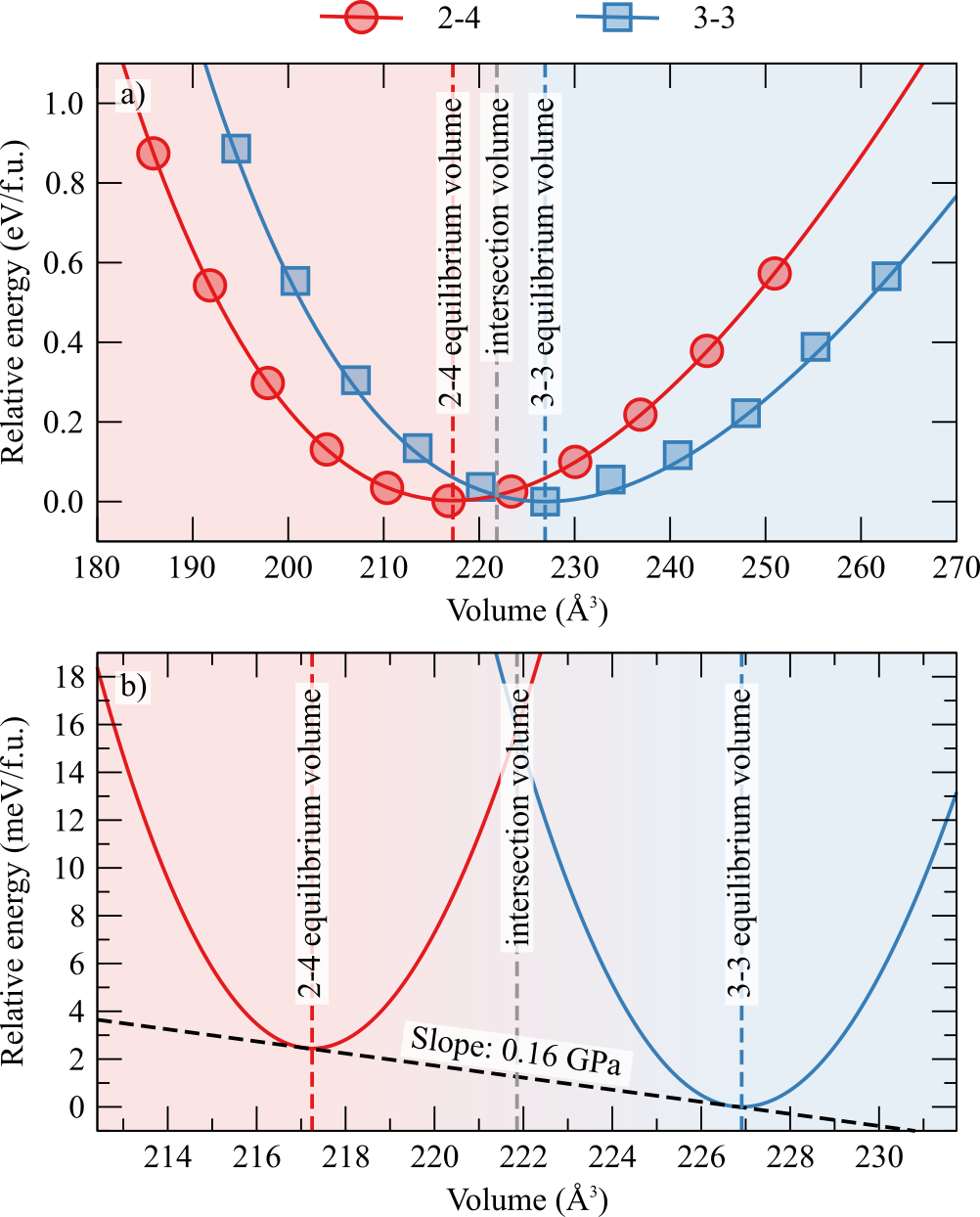}
\caption{\label{fig:pressure}a) Evolution of the energy relative to the 3-3 ground-state structure as a function of the cell volume. The data points are DFT results while the lines are fitted Murnaghan equation of state curves. b) Magnified view of the low energy region, showing also the common tangent and its slope. Dashed vertical lines indicate equilibrium and intersection volumes.}
\end{figure}

In a first series of calculations we compare EuMnO$_3$ in the two charge states as a function of the unit-cell volume. As we allow the cell shape to change while constraining the volume, this corresponds to the application of isostatic pressure. We see in Fig. \ref{fig:pressure}a that the 3-3 state (blue squares) has the lowest energy with a relaxed volume of 226.91 \AA$^3$, which compares well with the experimental\cite{Choithrani:2009ua} value of 228.75 \AA$^3$ and is stable for ambient, small positive and any (hypothetical) negative pressure. Our Murnaghan equation of state \cite{Murnaghan:1944ch} fits yield bulk moduli of 182 and 171 GPa for the 2-4 and 3-3 states respectively; the latter compares well with the 172 GPa measured in Ref. \onlinecite{Mota:2014dv}. We also see that, as expected from the ionic radii, the 2-4 phase becomes increasingly more stable at smaller volumes (positive pressure), having a relaxed volume of 217.25 \AA$^3$ with an energy only 2.44 meV per formula unit above that of the relaxed 3-3 phase. From the common tangent construction\footnote{The common tangent method exploits the fact that at the transition pressure, $P$, the Gibbs free energies, $G =E + PV - TS$, of the two phases (here $E$ is the internal energy, $V$ the volume, $T$ the temperature and $S$ the entropy) are equal. Then at the zero temperature of density functional theory, $E_1 + P V_1 = E_2 + P V_2$, (the subscripts indicate phase 1 and phase 2 respectively) so $P = -\frac{E_2 - E_1}{V_2 - V_1} = - \frac{dE}{dV}$.} shown in Fig. \ref{fig:pressure}b, we predict that the \textit{heteronuclear} charge disproportionated 2-4 phase becomes thermodynamically more stable than the 3-3 phase at pressures of 0.16 GPa and larger. Therefore at growth pressures above 0.16 GPa, the 2-4 phase should be thermodynamically favored over the 3-3. This pressure is lower than that previously predicted (2 GPa) for an AFM insulator to FM metal cross-over within the 3-3 phase \cite{Qiu:2017kl}, suggesting that three phases -- the insulating 3-3 AFM state, the metallic 3-3 FM phase and the 2-4 AFM phase, which we will show below is also insulating -- might coexist or compete in this pressure range. We note however that the relative energies are strongly dependent on the computational setup and in particular the choice of U$_\textrm{Mn}$ and that absolute values should be interpreted with care. 

\begin{figure}
\includegraphics[width=\columnwidth]{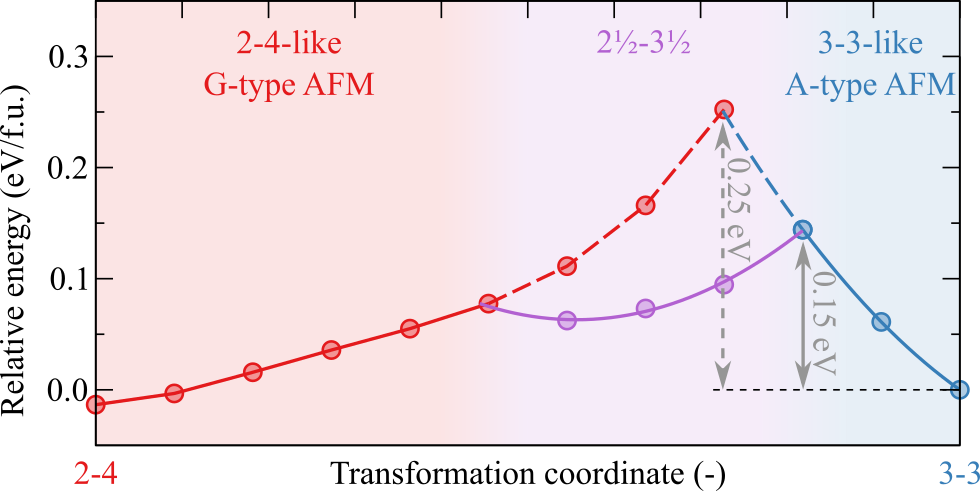}
\caption{\label{fig:barrier}Calculated energies of structures obtained by linear interpolation between the structures of the 2-4 phase (on the left) and the 3-3 phase (on the right) at the intersection volume of 221.23 \AA$^3$. Dashed portions of lines show metastable regions. Since the 3-3 and intermediate phases can only be stabilised over limited ranges of the transition pathway, we extrapolate their energies using a quadratic fit.}
\end{figure}

The question of the transition pressure required to \textit{transform} an existing 3-3 phase sample into the 2-4 phase is more difficult to answer, even aside from the ambiguities introduced by the choice of DFT exchange-correlation functional, since it depends on the kinetics and mechanism of the transformation. Ideally, an analysis would be based on a full nudged-elastic band (NEB) \cite{Henkelman:2000wb} calculation of the pathway between the relaxed 2-4 and 3-3 phases. We were unable, however, to achieve such a calculation, because of the competing magnetic and charge states at each image along the pathway. 
Instead, we linearly interpolate the lattice parameters and internal positions between the 2-4 and 3-3 phases at a volume of 221.23 \AA$^3$ close to the intersection volume, computing the energy of the two phases along the pathway as shown in Figure \ref{fig:barrier}. Compared to a full nudged-elastic band calculation this approach is expected to slightly overestimate the barriers. We note that compressing the 3-3 phase to the intersection volume that is the starting point for this calculation, corresponding to following the blue line in Fig.~\ref{fig:pressure}, requires a pressure of 4.11 GPa, which is already considerably larger than the value suggested by the common tangent construction. 

We show our results in Fig.~\ref{fig:barrier}. Interestingly, we find that for small deformations the 3-3 phase disproportionates into an intermediate (Eu$^{2+}$, Eu$^{3+}$)(Mn$^{4+}$, Mn$^{3+}$)O$_3$ phase, which we call $2\nicefrac{1}{2}$-$3\nicefrac{1}{2}$, in which half of the Eu and Mn ions have changed their oxidation state. We find multiple intermediate $2\nicefrac{1}{2}$-$3\nicefrac{1}{2}$ states, all of which are metallic, with different ion pairs changing their oxidation states and different magnetic orderings. Even when passing through this lower energy intermediate phase, the additional transition barrier at the intersection volume is $\sim$ 0.15 eV/f.u., which suggests substantial kinetic hindering of the transition at room temperature.

\subsection{Signatures of the phases and phase transition}

We will now extract signatures of the phases and the phase transition that could be monitored in the experimental studies that we hope to motivate with the present predictions. First, since the transition involves substantial changes in the crystal structure and lattice parameters, we expect it to be first order, with the corresponding divergence of the specific heat at the transition pressure. The difference in lattice parameters, shown in Table \ref{tab:lattice_mag}, while not large, should be detectable using standard x-ray diffraction. In particular, while the short axes are unequal in the 3-3 phase, we predict them to become almost equivalent in the 2-4 phase contracting by 0.06 and 0.21 \AA\ respectively during the transition. We note that we observe an unusual increase of the octahedral rotations with increasing volume, which we explain with changes in bonding as discussed in the appendix section B. The magnetic moments on the Eu sites represent a clear signature of the two phases: the $f^6$ Eu$^{3+}$ ion has zero total magnetic moment ($S=3$, $L=3$, $J=0$), whereas the $f^7$ Eu$^{2+}$ ion has a spin-only (since $L=0$) magnetic moment of around 7 $\mu_B$.  We also calculate a change in the spin-only Mn magnetic moment from close to the 4 $\mu_B$ expected for Mn$^{3+}$ in the 3-3 phase to close to 3 $\mu_B$ for Mn$^{4+}$ in the 2-4 phase.

\begin{table}
\caption{Lattice parameters and magnetic moments at the respective equilibrium volumes of the 2-4 and the 3-3 phase.}
\begin{ruledtabular}
\begin{tabular}{l | l l}
                   & 2-4   & 3-3   \\ \hline
a (\AA)            & 5.348 & 5.410 \\
b (\AA)            & 7.558 & 7.508 \\
c (\AA)            & 5.375 & 5.586 \\ \hline
$m_{Mn}$ ($\mu_B$) & 2.70  & 3.69  \\
\end{tabular}
\end{ruledtabular}
\label{tab:lattice_mag}
\end{table}

\begin{figure}
\includegraphics[width=\columnwidth]{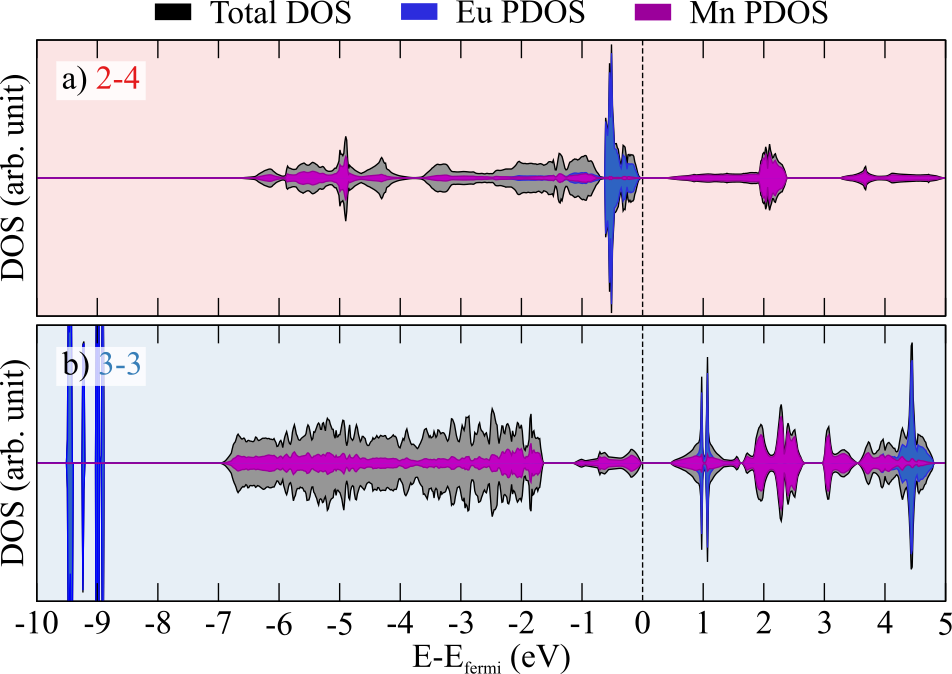}
\caption{\label{fig:dos}Element resolved electronic density of states (DOS) of a) the  2-4 and b) the 3-3 phase at their respective equilibrium volumes (217.25 \AA$^3$ and 226.91 \AA$^3$).}
\end{figure}

In Fig. \ref{fig:dos} we show the calculated electronic densities of states (DOS) of the two phases. We see that both phases are predicted to be insulating with DFT band gaps of 0.36 and 0.45 eV for the 2-4 and the 3-3 phases respectively. In the 3-3 phase, the occupied Eu majority $4f^6$ states are $\sim$2 eV below the bottom of the valence band, consistent with the XPS data \cite{Kurmaev:1998ic} and the band edge is composed of O$2p$ - Mn$3d$ hybridized states. The remaining majority Eu $4f$ state is empty and lies $\sim$1 eV above the Fermi energy in the conduction band. In the 2-4 phase, the filled Eu majority-spin $4f^7$ manifold forms the top of the largely O$2p$ - Mn$3d$ hybridized valence band, with the bottom of the conduction band deriving from empty Mn $3d$ states. The $4f^7$ dominated valence-band edge is consistent with XPS data for EuTiO$_3$ that also contains primarily Eu$^{2+}$ ions; these spectra also show small Eu$^{3+}$ contributions at 10 to 15 eV lower energies \cite{Hatabayashi:2009wk, Zhao:2012hi, Lv:2014ce}. The 2-4 phase thus has a Eu$_{f}$-Mn$_{e_g}$ gap whereas the 3-3 has a Mn$_{e_g}$-Mn$_{e_g}$/Eu$_{f}$ gap, and we anticipate strongly different transport behavior of electrons and holes in the two phases. Also given the completely different density of states, spectroscopic signatures for the two phases should show clear differences.

\begin{figure}
\includegraphics[width=\columnwidth]{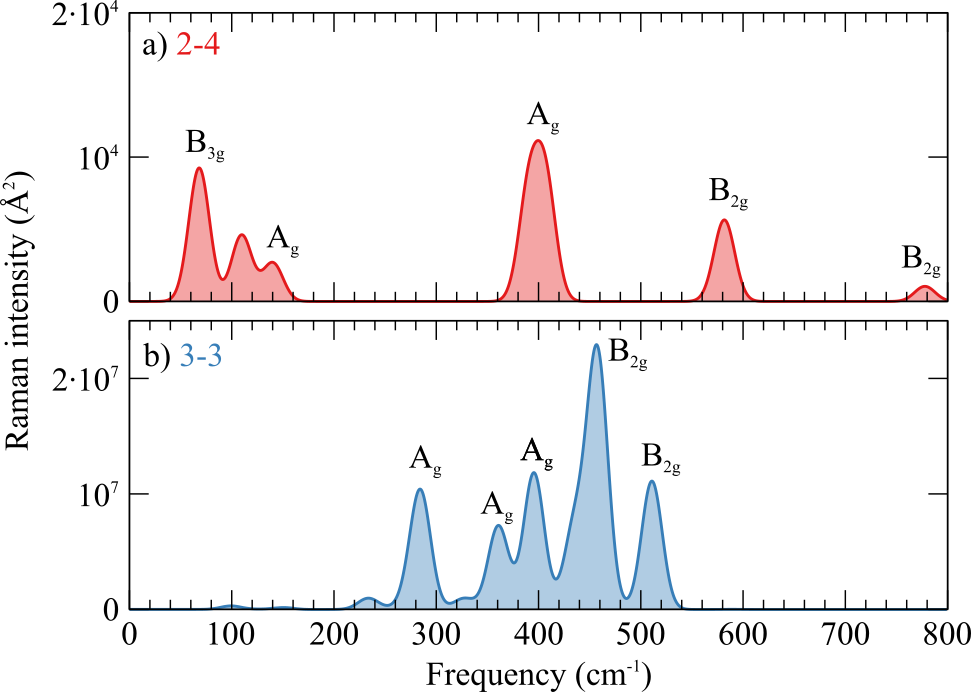}
\caption{\label{fig:raman}Computed Raman spectra for a) the 2-4 phase and b) the 3-3 phase at their respective equilibrium volumes. Due to the very different intensity axes, we provide frequencies and intenities for all Raman active modes in tables \ref{tab:raman24} and \ref{tab:raman33}. The labels show the irreducible representation of the dominant modes.} 
\end{figure}

In Figure \ref{fig:raman} we show the computed Raman spectra of the two phases at their respective equilibrium volumes. For the 2-4 phase shown in Figure \ref{fig:raman}a, we predict the presence of low frequency modes below 200 cm$^{-1}$ that correspond to octahedral rotations and deformation of the octahedra (see supplementary information for all eigenvectors), indicating that the 2-4 structure, while dynamically stable, has low-energy transitions towards the 3-3 phase. Besides the two Raman active A$_g$ (distortion) and B$_{2g}$ (2-out-1-in breathing) modes, the 2-4 phase also exhibits a very high frequency (3-out) breathing mode slightly below 800 cm$^{-1}$ (see Table \ref{tab:raman24} for more information). For the 3-3 phase shown in Figure \ref{fig:raman}b, all Raman active modes with high activities (see Table \ref{tab:raman33} for all activities) are concentrated in the range from 280 cm$^{-1}$ to 520 cm$^{-1}$. The lowest labeled A$_g$ mode corresponds to an octahedral rotation/distortion (see supplementary information for all eigenvectors), whereas the highest labeled B$_{2g}$ mode corresponds to an in-plane stretching. The intermediate modes are combinations of in-plane stretching with out-of-plane rotations. Experimental Raman investigations at ambient conditions \cite{Mota:2014dv} found an A$_g$ mode at 360 cm$^{-1}$ and a B$_{2g}$ mode at 610 cm$^{-1}$ along with an intermediate double peak that increased slightly in frequency with increasing pressure without any significant alteration of the overall shape of the spectrum up to 47 GPa. While our calculations for the 2-4 phase reproduce the peaks at 360 cm$^{-1}$ and 610 cm$^{-1}$ well (at 400 and 580 cm$^{-1}$ respectively), they do not show any intermediate Raman active modes. Conversely, the overall shape of the computed spectrum of the 3-3 phase is in better agreement with experiment but all the peaks are strongly shifted to lower energies by about 80 cm$^{-1}$. In view of this mismatch, we restrict our conclusion from our calculated Raman spectra to inferring that a transition from the 3-3 to the 2-4 phase should show up as a general reduction in the Raman intensity as well as the appearance of soft modes below 200 cm$^{-1}$.

Finally, there should be a marked change in reflectivity, resulting from the metallicity of the $2\nicefrac{1}{2}$-$3\nicefrac{1}{2}$ intermediate structure during the transition. 

\subsection{Epitaxial strain}

\begin{figure}
\includegraphics[width=\columnwidth]{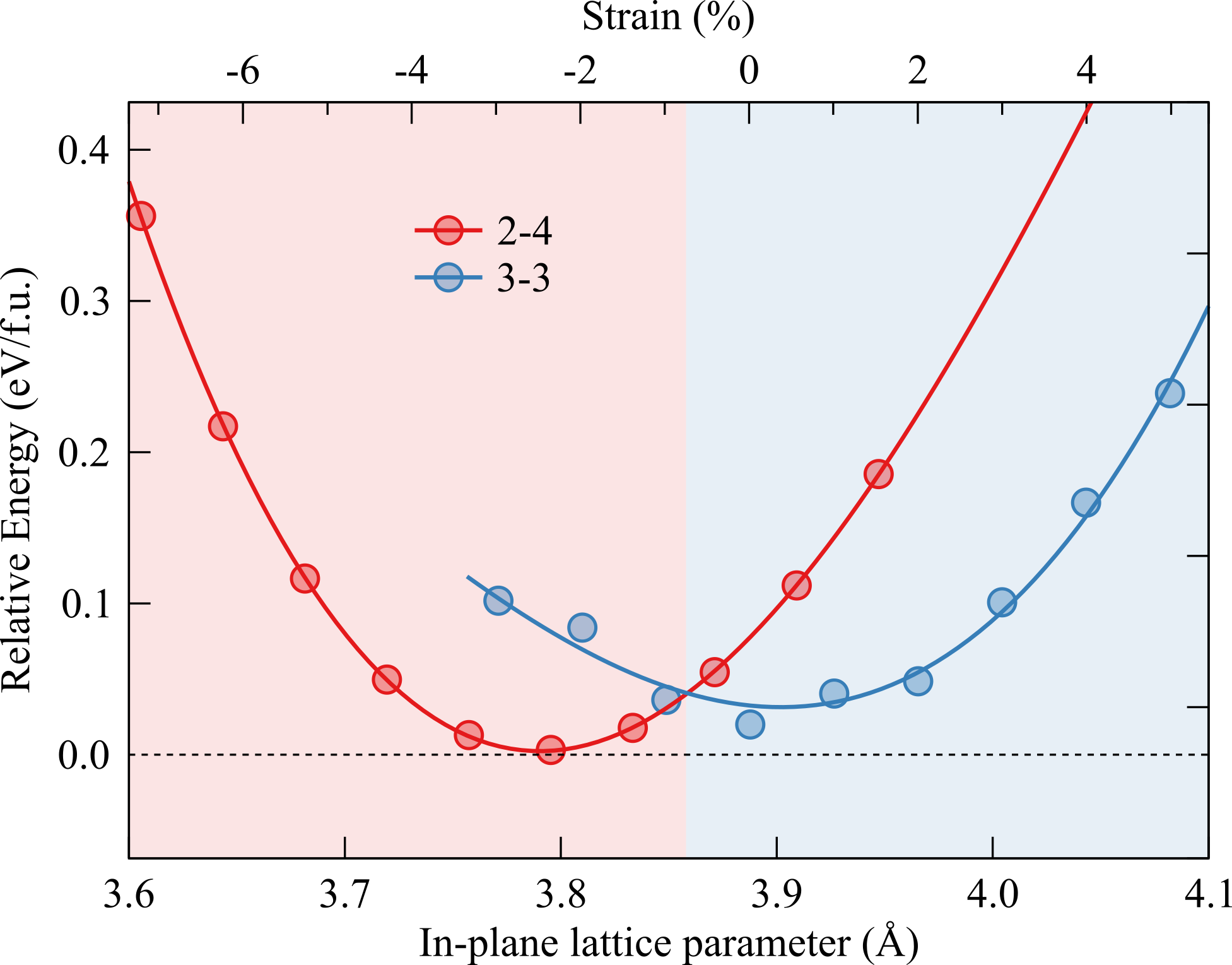}
\caption{\label{fig:strain}Energy per formula unit as a function of the in-plane lattice parameter (constrained to be equal and at 90$^\circ$ to each other) for the two phases. The zero of energy is the energy of the fully relaxed 3-3 phase, which is about 2 meV below the minimum of the strained 2-4 phase. The energy of the 3-3 phase at zero strain is about 20 meV higher than that of the fully relaxed 3-3 phase due to the constraint that the in-plane lattice constants be equal and perpendicular. The strain axis above the plot is defined relative to the average in-plane lattice constant of the relaxed 3-3 bulk phase.}
\end{figure}

Finally, we address the possibility of accessing the 2-4 phase using biaxial strain, provided for example by coherent heteroepitaxial growth on a substrate of different lattice constant. In Figure \ref{fig:strain} we show the energy relative to that of the fully relaxed 3-3 phase as a function the of in-plane lattice parameter. We impose the constraint that the in-plane lattice parameters are equal in length and perpendicular to each other to mimic the effect of heteroepitaxial growth on a cubic substrate. As a result of this epitaxial constraint, we find that the 2-4 phase, which has almost equal in-plane lattice parameters in its fully relaxed structure, has a lower minimum total energy than the 3-3 phase, with its strong Jahn-Teller distortions and resulting unequal lattice parameters. At the zero strain lattice parameter of 3.89 \AA, the 3-3 phase is lower in energy than the 2-4 phase, however for a small compressive strain of 1\%, which is routinely accessible with modern thin-film growth approaches, we observe a transition into the 2-4 phase. The non-quadratic behavior of the energy of the 3-3 phase under compressive strain reflects the difficulty of stabilising this phase under compression and the gradual transition into the $2\nicefrac{1}{2}$-$3\nicefrac{1}{2}$ intermediate phase. For a film grown on a substrate with lattice constant around 3.85 \AA, we might expect to achieve a coexistence of the two phases, providing an electronic analogue to the structural self-morphotropic phase boundary previously reported for strained BiFeO$_3$ \cite{Zeches:2009hk}. Such an electronic morphotropic phase boundary might have interesting divergences in susceptibilities associated with transport or reflectivity, in the same way that conventional morphotropic phase boundaries show divergent piezoelectric responses.

\section{Conclusions}

In summary, we have proposed the concept of heteronuclear charge disproportionation, in which two multivalent cations cooperate in inter-atomic charge transfer, and demonstrated its occurence theoretically in perovskite-structure EuMnO$_3$. We provided conditions of isostatic and biaxial strain at which such a heteronuclear charge disproportionation should occur, and propose that the transition should be experimentally accessible. The effect could compete with or even inhibit a previously predicted magnetically driven insulator-to-metal transition. Finally, to aid in the experimental elucidation of the pressure and biaxial strain-dependent phase diagram of EuMnO$_3$, we provided structural, magnetic and vibrational signatures of the transition.

The search for other materials exhibiting heteronuclear charge disproportionation should focus on materials with at least two multivalent cations on inequivalent sites. Possibilities could include mid-first-row transition metal perovskites with Ce or Eu A sites, such as  CeNiO$_3$, EuFeO$_3$ and EuCoO$_3$. In nickelates and cobaltites, the competition or cooperation with the other well-established instabilities could be a fertile ground for exploring new physics.

\section{Acknowledgements}

This work was financially supported by the ETH Z\"urich and by the ERC Advanced Grant program, No. 291151 as well as the SNF Professorship Grant PP00P2\_157615. CPU time was provided by the Euler cluster of the ETH Z\"urich, UBELIX (http://www.id.unibe.ch/hpc), the HPC cluster at the University of Bern and the Swiss Supercomputing Center (CSCS) under project s766. We thank Mael Guennou and Jens Kreisel for helpful discussions and feedback on the manuscript.

\bibliography{references.bib}

\newpage
\clearpage

\section*{Appendix}

\renewcommand\thefigure{A\arabic{figure}}
\setcounter{figure}{0}    
\renewcommand\thetable{A\arabic{table}}
\setcounter{table}{0}

\subsection{Effect of U on structure and magnetism}

\begin{figure}[h]
\includegraphics[width=\columnwidth]{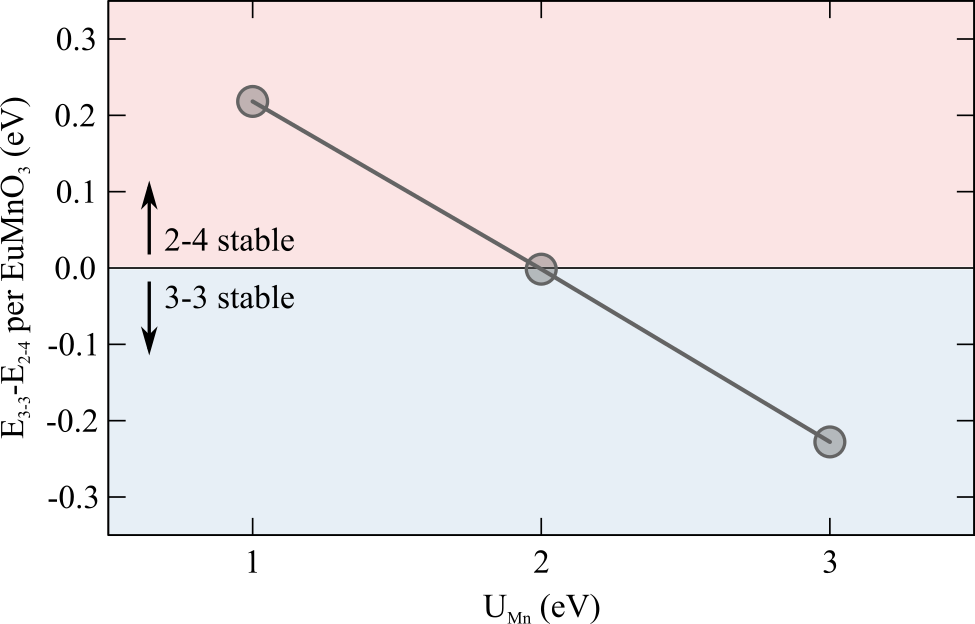}
\caption{\label{fig:struct_stab}Stability of the 2-4 and 3-3 phases as a function of U$_\textrm{Mn}$. Below U$_\textrm{Mn}\approx$ 2 eV, the 2-4 phase becomes more stable than the 3-3 phase.}
\end{figure}

In Figure \ref{fig:struct_stab} we show the relative structural stability of the 2-4 and 3-3 phases at their relaxed volumes and internal coordinates as a function of U$_\textrm{Mn}$. The experimentally observed 3-3 phase is only stable for U$_\textrm{Mn}$ > 2 eV, putting a lower bound for this parameter.

\begin{figure}[h]
\includegraphics[width=\columnwidth]{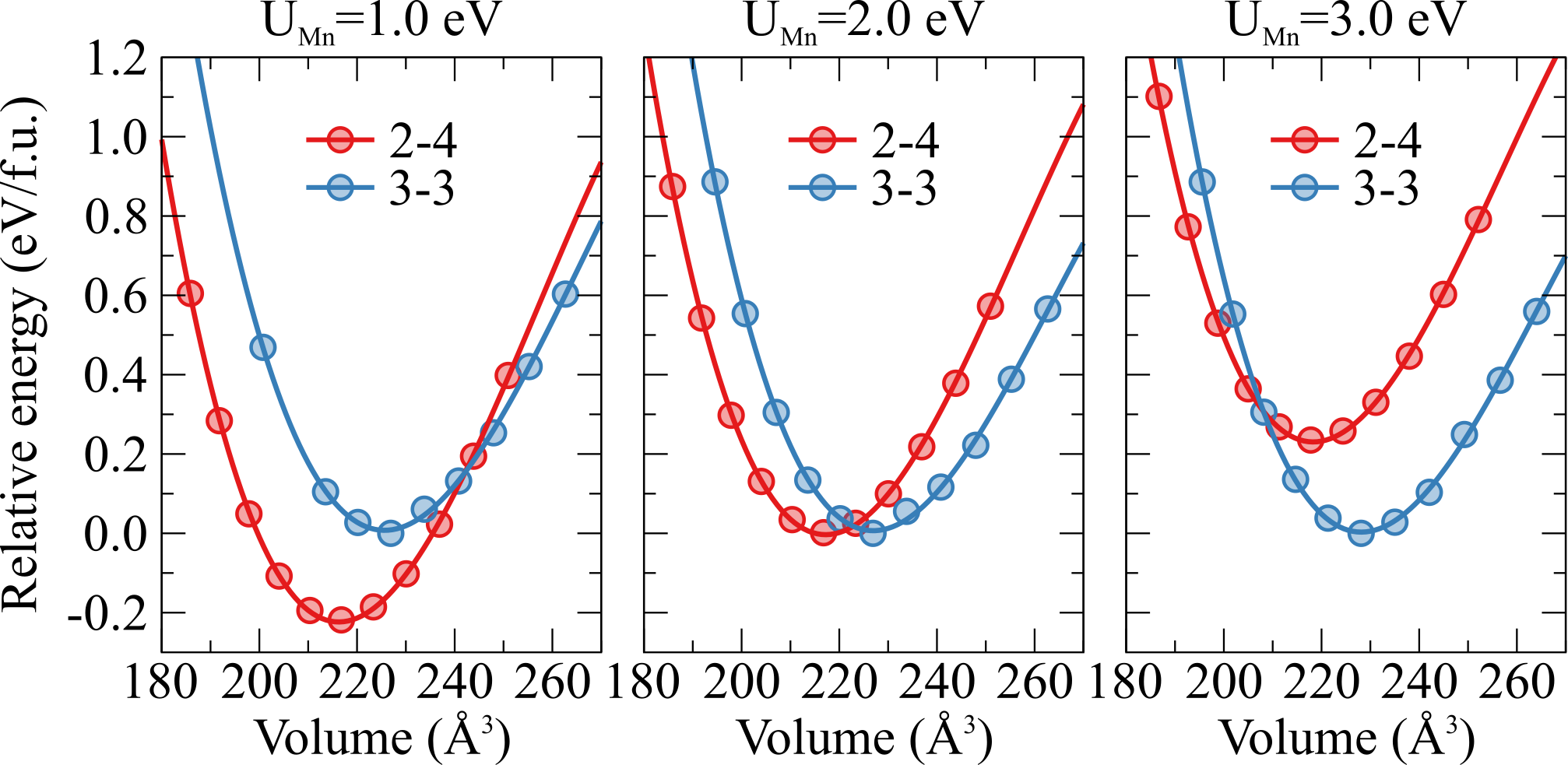}
\caption{\label{fig:ueffect}Effect of U on the relative stability of the 2-4 and 3-3 phases. In each case the zero of energy is set to the lowest energy of the 3-3 phase.}
\end{figure}

In Figure \ref{fig:ueffect}, we show the equation of state curves for U$_\textrm{Mn}$ of 1, 2 and 3 eV.

\begin{figure}[h]
\includegraphics[width=\columnwidth]{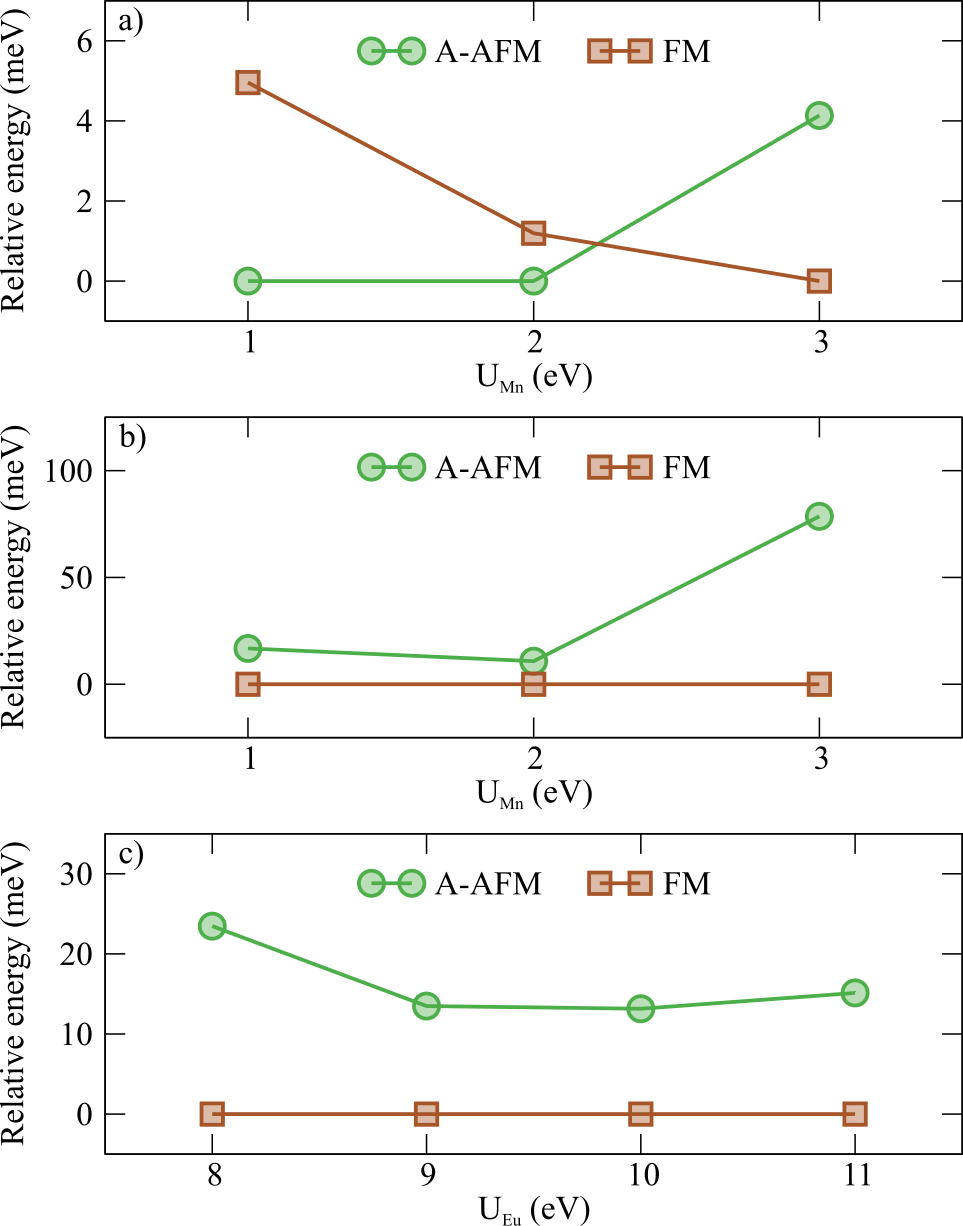}
\caption{\label{fig:mag_stab}Energies of the A-type AFM and FM phases relative to the respective ground state a) as a function of U$_\textrm{Mn}$ with Eu $f$ electrons in the core, b) as a function of U$_\textrm{Mn}$ with Eu $f$ electrons in the valence and U$_\textrm{Eu}$ = 10 eV and c) as a function of U$_\textrm{Eu}$ with a constant U$_\textrm{Mn}=2$ eV. In each case the zero of energy is set to the lowest energy magnetic phase.}
\end{figure}

In Figure \ref{fig:mag_stab} we show the relative energetic stability of the A-type antiferromegnatic (A-AFM) and the ferromagnetic (FM) magnetic phases with various computational setups. Figure \ref{fig:mag_stab}a is computed with Eu $f$ electrons in the core. As in a recent study \cite{Qiu:2017kl}, we find that at low U$_\textrm{Mn}$ values the A-AFM phase is more stable than the FM phase. This indicates an upper bound of $\sim$ 2.2 eV for U$_\textrm{Mn}$. When performing the same calculation for the PAW potential with Eu f states in the valence and U$_\textrm{Eu}$ = 10 eV we observe (Figure \ref{fig:mag_stab}b) that the FM phase is always more stable than the A-AFM phase, with U$_\textrm{Mn}$ $\sim$ 2 eV yielding the most similar energies, however still favouring the FM phase. As shown in Figure \ref{fig:mag_stab}c, U$_\textrm{Eu}$ does not improve the situation, the FM phase being stable for all values of U$_\textrm{Eu}$. For this reason, there is no combination of U values with the Eu $f$ electrons in the valence that correctly describes the charge state (2-4 vs. 3-3) as well as the ionic and magnetic structure. We choose U$_\textrm{Mn}$ = 2 eV and U$_\textrm{Eu}$ = 10 eV for the calculations in this work, while restricting the magnetic structure to AFM, as discussed in the main text.

\clearpage

\subsection{Pressure-dependent structural parameters}

\begin{figure}[!thb]
\includegraphics[width=\columnwidth]{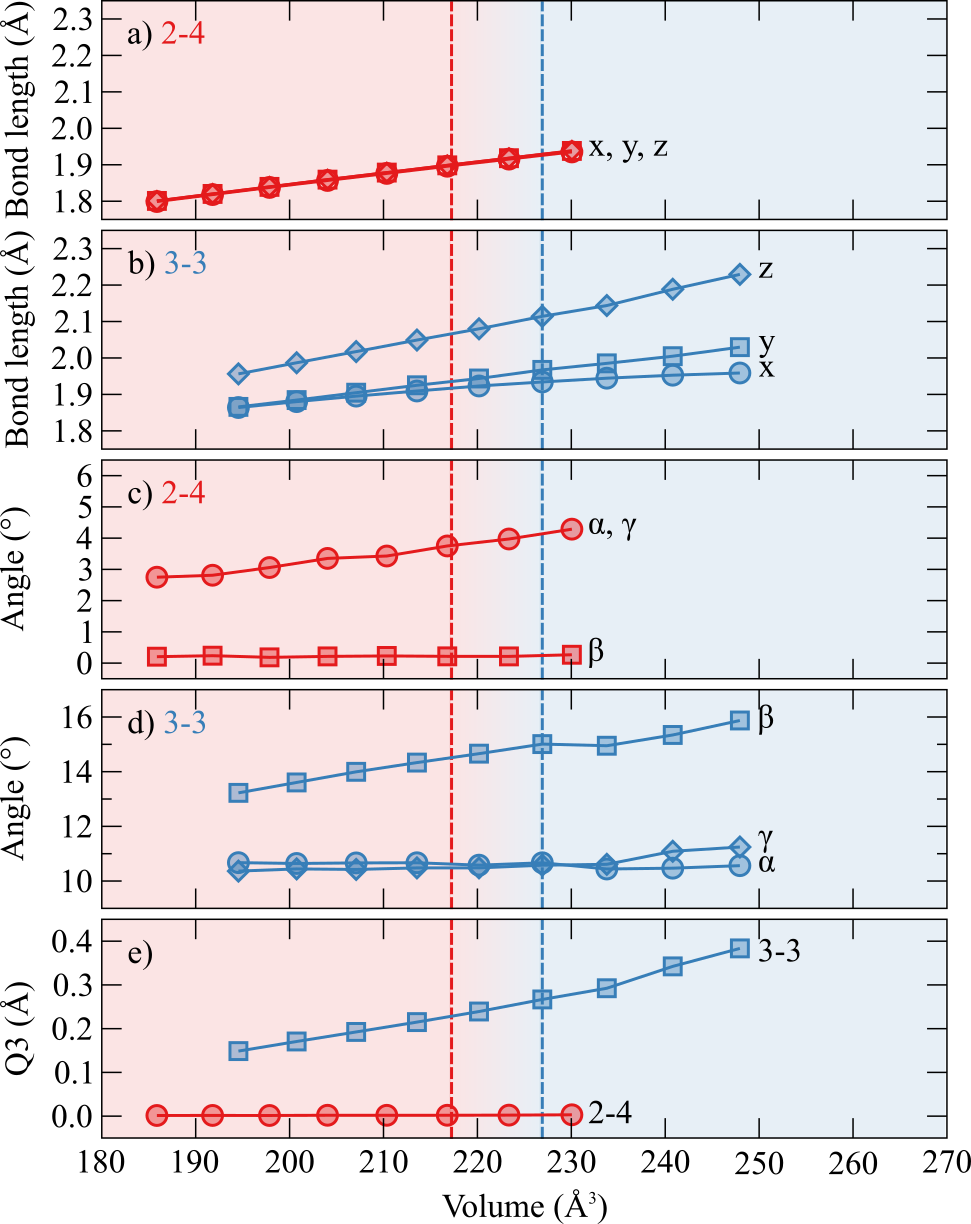}
\caption{\label{fig:ang_q3}Bond lengths in the a) 2-4 and b) 3-3 phase. The bond-lengths primarily along the x and z axes in the 3-3 phase switch in different octahedra due to the cooperative Jahn-Teller distortion. Octahedral rotation angles in c) the 2-4 and d) the 3-3 phase. Panel e) reports the Q$_3$ parameter characterising the magnitude of the Jahn-Teller distortion.}
\end{figure}

\begin{figure}[!thb]
\includegraphics[width=\columnwidth]{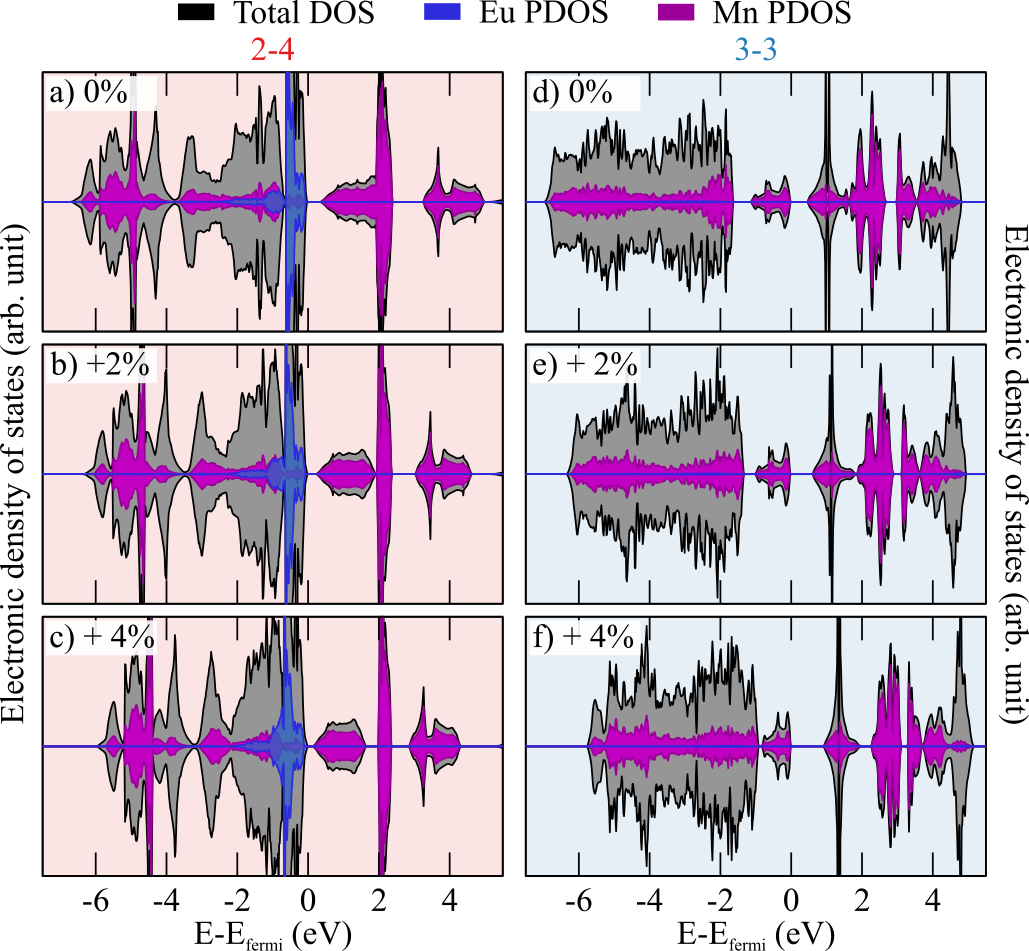}
\caption{\label{fig:pressure_dos}Volume dependent electronic density of states for a)-c) the 2-4 phase and d)-f) the 3-3 phase at the equilibrium volume as well as 2\% and 4\% expansion respectively.}
\end{figure}

In Figure \ref{fig:ang_q3} we report the bond lengths, octahedral rotation angles (determined via fitting Euler matrices to bring the octahedral bond vectors into coincidence with the crystal axes) and the Q$_3$ parameter characterising the magnitude of the Jahn-Teller distortion: 
\begin{equation}
\begin{split}
	Q_3 = \frac{1}{\sqrt{6}}(&2d_\mathrm{long, +} - 2d_\mathrm{long, -} \\
	&- d_\mathrm{short_1,+} +  d_\mathrm{short_1,-} \\
	&- d_\mathrm{short_2,+} +  d_\mathrm{short_2,-}) \quad .
\end{split}
\end{equation}
Here $d_\mathrm{bond,direction}$ refers to the length of a Mn-O bond pair (long, short$_1$ or short$_2$) along the positive (+) or negative (-) coordinate axis. 
As expected we see an increased bond length with increasing volume. Surprisingly, however, we see that the octahedral rotation angles increase in both phases with increasing pressure. This is opposite to the usual behaviour in perovskite oxides, where octahedral rotations are reduced or suppressed as the volume increases \cite{Rondinelli:2011jk}. More particularly it disagrees with the behaviour in epitaxial strained SrMnO$_3$ \cite{Ricca:2018} that is structurally similar to the 2-4 phase but where octahedral rotations with rotation axes perpendicular to elongated crystal axes disappear with increasing tensile strain. We can explain this difference in terms of the relative compressibilities of the Eu-O and Mn-O coordination spheres that result from the pressure-induced alterations of the valence-band density of states shown in Fig. \ref{fig:pressure_dos}. We see that increasing the volume has two effects on the DOS of the 2-4 phase. On one hand the $\sigma^*$ state just above E$_\textrm{fermi}$ shifts downwards, which is accompanied by an upwards shift of the respective $\sigma$ state in the valence band. At the same time the Eu $f$ states just below E$_\textrm{fermi}$ increasingly hybridize with the oxygen valence band states. The former leads to a weakening of the Mn-O bonds, while the latter increases Eu-O covalency and enhances octahedral rotations. For the 3-3 phase the changes are less clear, but we see a compression of the whole valence band, which implies that some Mn-O bonding states become destabilized, leading to longer Mn-O bonds. These changes in bonding lead to an enhancement of the octahedral rotations with increasing volume for the 2-4 and the absence of a decrease for the 3-3 phase. We also note that in the 3-3 phase we observe a small breaking of the \textit{Pnma} (a$^-$b$^+$a$^-$) symmetry with pressure, the out-of-phase $\alpha$ and $\gamma$ angles only being equivalent at the equilibrium volume. As expected, the 2-4 phase has no Jahn-Teller distortion and therefore a negligible Q$_3$, whereas the 3-3 phase has a Jahn-Teller distortion that increases in magnitude with increasing volume.

\clearpage

\subsection{Details on Raman-active modes}

In Tables \ref{tab:raman24} and \ref{tab:raman33}, we report the mode number, irreducible representation as well as the computed frequency and activity of all Raman-active modes in the 2-4 and 3-3 structure respectively. To complement this data we also provide v\_sim files with the eigenvectors in the electronic supporting information.

\onecolumngrid

\begin{table}[h]
\caption{Frequencies and activities of 2-4 Raman active modes (Imma) calculated in this work.}
\begin{ruledtabular}
\begin{tabular}{l | l r r r}
Mode & Irrep & Freq (THz) & Freq (cm$^{-1}$) & Raman Activity \\\hline
004  &   B1g &      1.953 &      65.145 &        1880.78 \\
005  &   B3g &      2.066 &      68.914 &        7469.62 \\
006  &    Ag &      3.264 &     108.875 &        2136.82 \\
007  &   B2g &      3.305 &     110.243 &        1025.72 \\
008  &   B3g &      3.318 &     110.677 &        1444.61 \\
009  &    Ag &      4.202 &     140.164 &        2665.71 \\
019  &   B3g &     11.699 &     390.237 &        6218.88 \\
023  &   B2g &     11.817 &     394.173 &        1482.53 \\
024  &    Ag &     12.213 &     407.382 &        7921.61 \\
028  &   B1g &     17.362 &     579.134 &        1046.15 \\
029  &   B2g &     17.472 &     582.803 &        4657.44 \\
030  &   B2g &     23.338 &     778.472 &        1045.68 \\
\end{tabular}
\end{ruledtabular}
\label{tab:raman24}
\end{table}

\begin{table}[!h]
\caption{Frequencies and activities of 3-3 Raman active modes (Pnma) calculated in this work.}
\begin{ruledtabular}
\begin{tabular}{l | l r r r}
Mode & Irrep & Freq (THz) & Freq (cm$^{-1}$) & Raman Activity \\\hline
005  &   B2g &      2.966 &      98.935 &      254689.88 \\
006  &    Ag &      3.047 &     101.637 &       44705.24 \\
009  &    Ag &      3.773 &     125.854 &       20267.71 \\
010  &   B3g &      3.896 &     129.957 &         807.49 \\
011  &   B1g &      4.447 &     148.336 &        3839.47 \\
012  &   B2g &      4.526 &     150.971 &      155946.39 \\
019  &   B1g &      6.133 &     204.575 &        6896.23 \\
020  &    Ag &      7.001 &     233.528 &      960937.06 \\
021  &   B3g &      7.415 &     247.338 &       50695.84 \\
024  &   B2g &      8.144 &     271.655 &      593883.78 \\
025  &   B1g &      8.229 &     274.490 &       16575.05 \\
027  &    Ag &      8.534 &     284.664 &    10152061.67 \\
031  &   B3g &      9.084 &     303.010 &       19555.94 \\
035  &   B2g &      9.816 &     327.427 &      964008.30 \\
038  &    Ag &     10.823 &     361.016 &     7249854.58 \\
042  &   B3g &     11.852 &     395.340 &       48894.35 \\
043  &    Ag &     11.865 &     395.774 &    11775456.04 \\
046  &   B1g &     12.553 &     418.723 &       68787.40 \\
047  &   B2g &     13.045 &     435.134 &     6813324.31 \\
048  &    Ag &     13.622 &     454.381 &     8695502.21 \\
050  &   B2g &     13.778 &     459.585 &    14376946.63 \\
052  &   B3g &     15.280 &     509.686 &      128934.98 \\
053  &   B2g &     15.328 &     511.287 &    10987955.23 \\
060  &   B1g &     17.677 &     589.641 &       14353.29
\end{tabular}
\end{ruledtabular}
\label{tab:raman33}
\end{table}

\end{document}